\documentstyle[preprint,aps]{revtex}
\def\be{\begin{equation}}
\def\ee{\end{equation}}
\def\bea{\begin{eqnarray}}
\def\eea{\end{eqnarray}}

\tighten
\begin{document}
\draft
 
\title{A Proposed Method for Measuring the Electric Dipole Moment \\
of the Neutron Using Acceleration in an Electric-Field Gradient  \\ 
and Ultracold Neutron Interferometry}

\author{M. S. Freedman,$^1$\cite{smfreedman} G. R. Ringo,$^1$ and T. W. Dombeck$^2$}

\address{$^1$Argonne National Laboratory, Argonne, IL  60439
\linebreak
$^2$Fermilab, Batavia, IL 60510}

\date{\today}
 
\maketitle
\bigskip
\bigskip

\begin{abstract}
The use of an ultracold neutron interferometer incorporating an 
electrostatic accelerator having a strong electric field gradient to 
accelerate neutrons by their possible electric moments is proposed as 
a method of measuring the neutron electric dipole moment.  Such 
electrical acceleration, followed by an amplifier and a generator 
of phase difference, could develop relatively large phase differences 
and these could be measured by a Mach-Zehnder interferometer.  This method 
might extend the sensitivity of the measurement by several orders of 
magnitude beyond the current limit of 10$^{-25}$ e.cm.  Furthermore the 
systematic errors in such a measurement could be significantly 
different from those of the current EDM experiments.
\end{abstract}

\bigskip
\bigskip
\bigskip


\newpage
\tableofcontents

\newpage
\section{Introduction}
The existence of a neutron electric dipole moment (EDM) would demonstrate the 
violation of time reversal invariance.  The measurement of the EDM is one of the 
most sensitive time reversal tests available.   
The experimental searches for the EDM by the method of Norman Ramsey and 
his colleagues [1,2] have increased in sensitivity by a factor of about 
10$^5$ beyond their first measurement which was itself quite a sensitive test.  
Most measurements of the neutron EDM have been NMR measurements [1,2,3] in
which one looks for the effect of a strong electric field on the precession 
rate of the neutron spin.  To enhance sensitivity all such recent measurements 
are made with ultracold neutrons (UCN; T$<$3mK) which can precess for long 
times while stored in containers made of materials which reflect them 
coherently, even at normal incidence.  Only one other measurement, by 
Shull and Nathans [4], was not done by an NMR measurement.  In 1967 they 
set a limit of about 7 $\times$ 1$^{-22}$ e.cm for the EDM (expressed in the usual 
fashion of the electron charge times the separation of electrons of opposite 
charge, needed to give the measured moment).  They looked for the effect 
of the strong electric field gradients in atoms on the EDM via the imaginary 
term in the neutron scattering amplitude of cadmium.  However, the best current 
measurements [1,2] using the precession method give (95\% CL) $\leq$ 10$^{-25}$ e.cm 
for the EDM which is a factor of thousands lower than the Shull and Nathans experiment.  
In view of the importance of the EDM and the great delicacy of the measurements, 
it seems very desirable to have a test with different systematic errors and a 
sensitivity comparable to the Ramsey method.  This paper describes a method which 
is not only different from the Ramsey method but could possibly provide higher 
sensitivity.
  
Measurement of the EDM by neutron interferometry was proposed by Anandan [5] in 1982.  
His method was to measure the uniformly accumulating phase shift of a polarized neutron 
placed in a strong uniform electric field along the spin axis.  He did not claim 
significant gains in sensitivity over the precession method in which the neutron 
precesses in a plane orthogonal to the electric field.  We are proposing a different 
interferometric method [6] which should give a considerable advance in sensitivity.  
No scheme can violate the time-energy uncertainty principle which limits 
$\Delta \epsilon \geq h/(\Delta t)$
for one neutron.  The $\Delta \epsilon$ limit for $\Delta t$ of the order of the neutron 
lifetime is currently 
approached by the precession method for an EDM limit of order 
10$^{-27}$ e.cm, but for our proposal the same $\Delta \epsilon$ corresponds to an 
EDM smaller by about 10 orders of magnitude (see Sect.\ II).
	Several proposals are the main themes of this paper.  These are:

\begin{enumerate}
\item	The amplitude of each UCN wave function (velocity $\leq$ 4.5 m/s 
for us; $\lambda \sim$ 100 nm) would be split into two overlapping partial waves 
with spins parallel and antiparallel to a magnetic guide field.  This scheme 
would maintain the partial waves coherence for long times while suppressing 
adventitious phase shifts.  These partial waves are the ``arms'' of our interferometer.  
For this purpose we propose the use of UCN with their spins orthogonal to the 
magnetic guide field direction.  This is described in Sect.\ II-A.

\item We would generate a velocity difference and thus a desirable (microscopic) 
spatial separation between the partial waves by a differential acceleration of their 
EDMs in an electric field gradient for long times, while retaining a significant range 
of UCN velocities in the electric gradient.  These are accomplished in the
``Open Box Accelerator'' with a dihedral ceiling, Sect.\ II-B.

\item We suggest two methods for converting the spatial separation into a greatly 
enhanced potential difference.  Both produce a growing phase difference between 
the partial waves while conserving the UCN fluence.  They are described in Sect.\ II-C, 
``Phase Difference Generation'' and in Sect.\ II-D, ``The Assembly'' incorporating 
a drift space.

\item Finally, the phase difference is measured by a Mach-Zehnder interferometer
in which the first reflector is a magnitized mirror which sends one spin state in one
path and the other in the other path.  One of the spin states is rotated $\pi$ radians and
they are put together at the recombiner. 
\end{enumerate}

We also present fairly detailed analyses of some significant sources of 
systematic error; in Sec. IV-A, ``${\bf v} \times {\bf E}$''; 
in Sect.\ IV-B, 
``Magnetic Fields''; in Sect.\ IV-F, ``Gravitational and Sagnac Effect''; and 
in Sect.\ IV-H, ``Control, Calibration and Alignment Measurements''.

This proposal is clearly an optimistic extrapolation of a barely started 
technology which uses several untested ideas, in fact there has been significant 
theoretical criticism of our basic assumption, item 3 above.  No one should
embark on such a measurement without testing our assumptions on a small scale.  
In Sect.\ V we suggest relatively cheap ways of testing by measuring the well-known 
neutron magnetic moment using our proposed method.

The aim of this paper is to offer suggestions and not to make judgments as 
to matters of strategy, let alone engineering.  All our numbers are to be 
taken as illustrative.  

\section{The Proposed Instrument}
\subsection{Principle of Operation}
Splitting the wave function amplitude of each particle in the beam is the first step 
in interferometers.  The common way of doing this is by separating the partial waves 
geometrically.  For us this would pose very difficult construction problems.  Owing 
to the low phase space density of the UCN sources, we need to obtain UCN from broad, 
uncollimated polychromatic sources.  We need to contain these neutrons for perhaps 
10 minutes in apparatus of the order of 1 meter in size.  We would have to make the 
two separated paths for each UCN identical to a few nm although the paths are multiply 
folded and kilometers long.  In addition, slight differences in the vibration histories 
of the separate containers might be fatal.

A better way to separate the paths would be by their quantum states.  This 
is relatively easy to arrange.  The beautiful atomic interferometer measurement 
of the acceleration of gravity done by Kasevich and Chu [7] demonstrates that 
this technique is sound in principle and is practical.  In our EDM scheme the spatial 
separation of the two states is of the order of nanometers at most and they are 
reflected simultaneously from nearly the same area at all times.  It is not necessary, 
of course, to match the varied path lengths of different UCN.
  
All our computer modeling and our use of semi-classical equations of motion refer 
to point-like particles.  The actual interactions of UCN with fields and media 
surfaces extend over regions at least the size of the UCN wavelength.  There is 
ample precedent for the application of semiclassical methods in following the 
course of development of the quantum mechanical properties of ensembles of particles, 
e.g.\ phase difference, in such billiard-like problems as ours [8].
	
We first consider a simple accelerator followed by a detector (Fig.\ 1) 
to introduce the main ideas of the scheme.  We start with a pulsed beam of 
monochromatic UCN polarized in the direction they are moving (the x-direction) 
with a weak magnetic guide field in the x-direction.   Applying an RF-field, at 
the Larmor frequency and normal to x, for a suitable time will rotate the spin to 
a direction normal to the x axis and it will start to precess around that axis.  
This state is equivalent to coherently superposed eigenstates of the magnetic field, 
one with spin parallel to x, the other anti-parallel.  These constitute the two 
partial waves of our interferometer.  We call this pair, ``bipolarized''.  This 
superposition has been demonstrated both experimentally and theoretically by 
Summhammer $\it et\,al.$ [9] and by Badurek $\it et\,al.$ [10].  In particular 
they showed that the separated coherent bipolarized neutron eigenstates, when 
recombined interferometrically, summed their independent phase shifts.  These 
phase shifts correspond to polarization rotations, even from interactions which
were not spin dependent.

The UCN flow through an electric field gradient in the x-direction for a time.  
There one spin state is accelerated by the force of the gradient acting on one 
electric dipole but the other state, of opposite spin and dipole moment, is decelerated.  
This leads to a growth of a phase difference, $\Delta f$, between the spin states.  The UCN 
next pass into an RF coil parallel to the first coil which is then turned on at the 
same frequency and phase as the first RF coil.  At this time, common to both spin states, 
the bipolarized state and its differential phase gain are terminated by applying that 
field for the same duration as the first.  Thus the interference of the two spin states 
occurs after traversing equal-time arms as in Kasevich and Chu [7] and as in the Ramsey 
precession method [1,2,3] instead of equal-space length arms as in Mach-Zehnder particle 
interferometry.
	
The $\pi$/2 spin rotation induced by the second RF leaves it anti-parallel with the initial 
spin if there is no EDM and slightly off that if there is a polarization change dependent 
on the phase difference generated in the field gradient.  (In practice the polarization 
change is more sensitive to the phase difference of the UCN states if the second RF is 
shifted by $\pi$/2 in phase relative to the first.)  The polarization would be detected by 
standard means after the second RF field.  All this is very similar to the precession 
measurement [1].
	
For the phase difference $\Delta \phi$ between the spin states generated in $ta$ seconds in 
the accelerator, we follow the development of Greenberger and Overhauser [11].  
It is given by the line integral

\begin{equation}
\hbar \Delta \phi = \int^{t_a}_0 \Big( V_{\rightarrow} - V_{\leftarrow} \Big)
dt = \int ^{t_a}_0 \Delta Vdt
\end{equation}
where the potential energy difference between the states, labeled by the arrows, 
is a small perturbation on the free particle Hamiltonian.  Since the fields 
between the RF spin flipper, the electric, magnetic and gravitational, are all 
static, total energy is conserved,

\begin{equation}
\Delta W = \Delta V + \Delta T = 0 \,.
\end{equation}
$\Delta T$ is the kinetic energy difference.  Hence

\begin{equation}
		\hbar \Delta \phi = \int ^{t_a}_0 - \Delta T dt \,.
\end{equation}
For small $\Delta v_x$ 

\begin{equation}
		- \Delta T = m v_x \Delta v _x
\end{equation}
where $\Delta v_x$ is the difference in velocity of the two spin states produced by the 
electric field gradient, and m is the neutron mass.  (Here we are ignoring the 
relatively large guide field interaction with the magnetic moment, 2$\mu _mB$, because 
it could be canceled, for example, by reversing the magnetic field non-adiabatically 
in mid course, or by other ways discussed in Sect.\ IV-B.)  The force of the electric 
gradient on the dipole gives after $t_a$ seconds

\begin{equation}
\Delta v_x = {2\mu _e t_a \over m} \Bigg( {\partial E_x \over \partial x} \Bigg) \,.
\end{equation}
Here the factor 2 accounts for two spin states, $\mu _e$ is the neutron EDM (arbitrarily 
assumed to be parallel to the spin vector) and $E_x$ is the x-component of the electric 
field.  We then have from (3), (4), and (5)
		 		
\begin{equation}
\hbar \Delta \phi = \mu _e t^2_a \Bigg( {\partial E_x \over \partial x} \Bigg) v_x \,.
\end{equation}
In the precession method

\begin{equation}
\hbar \Delta \phi = 2\mu _e E_x t_P \,,
\end{equation}
where $t_P$ is the time spent in the electric field.  The interferometer in Fig.\ 1 would, 
in practice, be far less sensitive than Refs.\ [1] and [3].  Monochromatization and 
pulsing would greatly reduce the number of UCN.  Also the time in the gradient is 
very small compared to the times, $t_P$, of stored UCN.  The above calculation makes the 
important point that the effect of the gradient is proportional to $t_a^2$.
	
There are several ways of realizing a UCN interferometer with a large $t_a$.  We will 
describe two, each consisting of an accelerator with a special UCN storage arrangement 
in order to gain the largest possible $\Delta x$, followed by an amplifier and a 
generator of 
phase difference.  The latter are shown dashed in Fig.\ 1. 

\subsection{The Proposed Open-Box Accelerator}
To hold the full UCN velocity range for a long time in the accelerator one might consider 
putting the transversely polarized UCN into a rectangular box of totally-reflecting optical 
flats, made of insulators, located in a static, uniform, longitudinal, electric-field 
gradient.  For a given longitudinally polarized partial wave, after each spectral 
reflection from the walls perpendicular to the x direction of the polarization, 
the x-component of the velocity reverses relative to the acceleration which is 
unidirectional; it is well established that the polarization is conserved on reflection [2] 
and that UCN reflection is specular [12].  It is obvious that the reversal of the signs of
$v_{x\rightarrow}$ and $v_{x\leftarrow}$ 
with preservation of their magnitudes on reflection at walls normal to the x-axis entails 
the reversal of their difference $\Delta v_x = v_{x\rightarrow} - v_{x\leftarrow}$.  
Thus the $\Delta v_x$ gained in each traversal effectively 
cancels the $\Delta v_x$ gained in the previous traversal of the box. 
	
We considered a number of accelerator schemes that do not suffer such cancellation of 
$\Delta v_x$.  By far the most favorable in overall system simplicity, size, cost, 
sensitivity, and minimization of systematic errors is the ``open box'' shown in Fig.\ 2.  
We propose to do the x-direction differential acceleration in an open box made of 
dielectric UCN-reflecting plates, say, 50 cm long in the y direction, by 10 cm (x) as 
floor; and having 2 ceiling plates each 5 cm wide in the x direction, joined at a 
small dihedral angle ($\sim$0.01 radians) in the z direction, peaked in the middle (x = 0).  
The box would be 5 cm high.  Two 5 by 10 cm wall plates would close the y ends.  
The box would be open in the $\pm$x directions to avoid wall reflections of $v_x$.  Provided 
$v_z/v_x \sim$ 10 or greater it is possible to reflect UCN of sufficiently broad, 
low- x-velocity range back and forth along the x coordinate without suffering the 
reversal of $\Delta v_x$ indicated above.  This is accomplished by several reflections 
from the 
tilted top plates in a manner described in the Appendix.  This arrangement retains in 
the steep electric gradient region many more UCN (by 10$^3$ for $t_a$ = 100 s) than would 
remain if the ceiling were flat.

A plausible, qualitative description of the processes in the open and closed boxes follows.  
Consider the bipolarized state being reflected back and forth from the dihedral ceiling and 
planar floor and moving towards $\pm$x, but without the electric field.  Clearly both 
partial waves will reverse their x motion at exactly the same $|$x$|$ value, on the ceiling.  
Now with the electric field turned on its gradient will ``pull'' one polarization state 
towards the + x direction to which it was aiming, and repel the other state away from 
that end, thus helping it turn around.  Obviously, the latter (\underline{lagging} 
wave $\bf v_{\leftarrow}$, in Fig.\ 9 
of the Appendix) will reverse its x motion at smaller $|$x$|$ than the former.  After the x 
motion reversal the leading and lagging waves interchange roles.  Thus their 
$\Delta v_x$ always preserves its sense with respect to the acceleration direction, 
i.e., $v_{\rightarrow} = v_{\leftarrow} + \Delta v$.  
In the case of the closed box, on the contrary, the \underline{leading} wave reverses 
first on every x-reflection, thus reversing the sign of $\Delta v_x$, leading to its near 
cancellation.  See the Appendix for further remarks. 
	
In the horizontal direction UCN with a $|v_x|$ of 0.3 m/s in the center of the open box 
can be retained effectively with an optimal dihedral ceiling angle of about $|$0.006$|$
radians from the x axis according to our computer simulations.  A complication is 
that most UCN with $v_z <$ 3 m/s will not behave properly in the open box and will be lost.  
About 1\% of all the UCN with $v <$ 4.5 m/s can be retained for 100 seconds.  The rest 
quickly escape and are absorbed on the vacuum walls.

Note that a shallow concave roof, either parabolic or circular, with an average 
slope matching that of the dihedral roof, $|$0.006$|$ rad, is even more effective in 
retaining low $v_x$ UCN.  However such concave curvatures counteract the electric 
differential acceleration (see Sect.\ II-C, ``Convex Floor Amplifier'') so 
they cannot be used.
	
A static, hence energy conservative, electric field gradient (Fig.\ 3) would be 
supplied by four horizontal rod electrodes at least ~0.6 m long in the y direction, 
arranged in a symmetrical diamond array; the pair in the yz plane opposite in 
polarity to the pair in the xy plane, the pair members spaced 50 cm apart (Fig.\ 2).  
(The electrodes in the xy plane, connected in a loop, could serve as the coil of an 
RF spin rotator for both the accelerator box and the drift space.)  The gradient 
($\partial E_x/\partial x$) varies only a few percent within the box along any direction.  
Along the x 
centerline plane of the box the electric field is zero (Fig.\ 3).   Hence the 
electric polarization of the neutron induced by the weak field in the box is 
negligible, in view of the neutron's very small polarizability [13] (0.9 $\times$ 
10$^{-31}$ e.cm 
in fields of 10$^6$ V/m).  Moreover, induced polarization cannot contribute to the 
differential acceleration.  The dihedral ceiling and the floor, in practice, 
would be much wider in the x direction than 10 cm to minimize dielectric edge 
effects on field smoothness in the middle 10 cm as explained in Sect.\ IV-A.    
The dihedral angle should extend only to $|x|$ = 5 cm, thus confining about 1/3 of 
the UCN with $|v_x | \leq$ 0.3 m/s (at their entrance into the box) to this range.  
The dielectric mirrors should be optically smooth and polished but could 
tolerate surface waviness and misalignment at the one tenth milliradian level.
	
For each component of the bipolarized state the electric acceleration is nearly 
constant, unidirectional, independent of the sign of the small $v_x$ and opposite that 
of the other spin component, and energy conservative.  The orientation of $\bf\Delta v$ 
is nearly constant so $\Delta v_x$ grows as $t_a$.  The resulting spatial separations 
between the members of the partial wave pairs as they leave the accelerator, 
$\Delta x_a$, are nearly the same, within a few percent for all partial wave pairs, 
even though their $<\Delta \phi>$ = 0.  On every traversal $v_x$, hence $\Delta T$, hence 
$\Delta \phi$, all 
reverse, hence cancel; Eqs.\ (3, 4).  Any variation in $\Delta x_a$ would be due mainly 
to the small variation in the magnitude of the electric field x-gradient over 
the box which the UCN scan.  It is the separation, $\Delta x_a$, of the partial wave 
packets rather than the velocity difference $\Delta v_a$ from the accelerator, which 
leads in the converter-amplifier (Sect.\ II-C), to a large gain in average 
gravitational potential difference, which difference persists in the drift space.  
	
Gravity causes all the trajectories in the accelerator and amplifier to be 
sections of vertical parabolas; the partial waves bounce together along the
x and y coordinates, the pair members separating very slightly in x with time.  
There will be a comparatively negligible z separation and no y separation.  
Gravity was included in all computer simulations.

\subsection{Phase Difference Generation}
\underline{The Converter}.  Since $\Delta \phi$ is negligible in the open box accelerator 
it is necessary to convert $\Delta x_a$ to a potential energy difference which generates 
phase in the drift space.  This is accomplished by what we call the gravitational 
amplifier; Fig.\ 2.  Here after $t_a$ seconds of electrical acceleration a 10 cm 
y-length section of the floor of the accelerator is tilted 0.01 rad clockwise 
so the UCN drift to the right past the edge of a ceiling slot.  There they rise 
into a drift space, in the process hitting a 45$^{\circ}$ mirror.  The mirror converts 
$\Delta x_a$ to $\Delta z$, an extra gravitational deceleration distance for one spin state, 
always the same one.  That in turn produces a $\Delta v_x$ = $g\Delta z_{initial}/v_x$,
$\sim$ 100 
times the velocity difference in the accelerator, and a gravitational potential 
energy difference $\Delta V = mg\Delta z$ between the spin states which will lead to a 
detectable phase difference in the drift space [Eq.\ (1)].  Note that the $\Delta V$, 
proportional to $\mu _e$, arises from the gravitational field acting on the ``large'' 
neutron mass after an electric field gradient acts on the tiny electric dipole.  
This increase in $\Delta V$ by $\sim$ 10$^3$ accounts for the increased rate of phase 
gain after the mirror.
	
The drift space for practical reasons can only be 10 cm or so long in x 
but x-reflection on closed gates or fixed walls does not reverse the phase 
gain, unlike the case in the open box where $\Delta v_x$ does not reverse when $v_x$ does.  
This phase gain will be continuous and proportional to the time $t_d$ the UCN is 
in the drift space.  As in the case of the simple interferometer of Sect.\ II-A, 
the field, in this case gravity, is conservative so Eq. (1) 
$[\hbar \Delta \phi = -\int^{t_a}_0 \Delta V dt]$ 
applies with limits those of $t_d$.  Because of the parabolic paths $\Delta V$ fluctuates, 
even in sign when the (z,t) trajectories cross, but $\Delta \phi$ gains without changing 
sign, as shown in Fig.\ 4.
	
Since our fields are conservative,
\begin{equation}
-\Delta T = \Delta V = mg\Delta z \,.		
\end{equation}
At the 45$^{\circ}$ mirror	

\begin{equation}
		\Delta z (initial) = \Delta x_a \,.	
\end{equation}
Integrating Eq.\ (5) over time we have

\begin{equation}
		\Delta x_a = (\mu _e t^2_a /m ) (\partial E _x / \partial x ) \,.	
\end{equation}
Using (8), (9), and (10), and doing the integral in Eq.\ (1) over the time 
interval $t_d$ common to all the neutrons, we get

\begin{equation}
		\hbar \Delta \phi = 0.67 g\mu _e t_d t^2_a (\partial E_x /\partial x ) \,.
\end{equation}
The factor of 0.67 comes from averaging the varying $\Delta z$ over the many 
parabolic bounces of each UCN.  A further factor of 2 arises from taking 
the difference between $\Delta \phi$ of Eq.\ (11) and the $\Delta \phi$ from another set of 
measurements with the electric field reversed giving an effective $\Delta \phi$

\begin{equation}
	\hbar \Delta \phi _e = 1.3 g\mu _e t_d t^2_a (\partial E_x /\partial x ) =  \,.	
\end{equation}
Note that Eq.\ (12) is independent of v.  This is a ``white fringe'' interferometer.  
The electric field reversal is very important because it automatically helps 
to take care of a number of possible sources of error such as a slight error in 
the timing of the second spin flip or an adventitious magnetic field gradient (Sect.\ IV-B).
	
As noted in the introduction there have been objections raised that there is no 
generation of $\Delta \phi$ in the drift space in our experiment.  There is contrary 
evidence, 
however, from the Kasevich and Chu [7] experiment which produces a phase shift.  
Their experiment is like ours in that the potential energy difference is developed 
along strongly overlapping paths of the two partial waves.   Also the Hamiltonian 
is formally almost identical to ours, $-\hbar ^2 (\Delta ^2 /2m) + mgz + \mu _m B$, 
where $B$ is the external 
guide field in our case and the internal magnetic field of the sodium atom electrons 
at the nucleus in Ref.\ [7].  It is also like ours in that it is energy conservative 
in the periods when phase differences are being generated during the drift although 
not conservative when the velocity separation of the partial waves is being produced 
by the exciting laser pulses.  They differ in that their trajectories are linear 
along g, whereas our trajectories follow parabolas in the xz-plane.  
Nevertheless we do not expect any adverse consequence of this multi-dimensionality 
since we have also done a calculation using x and z velocity terms which agrees with 
the phase from Eq.\ (1).  (Proposed simplified tests of these and other questions 
are described in section V.)
	
\underline{The Convex Floor Amplifier (CFA)}.  
The second type of amplifier is somewhat more 
complicated but has the advantage of much greater sensitivity.  It may well be 
that systematic errors will make this advantage academic.  However, one cannot 
be certain of this so we believe it is worth considering.  In this amplifier 
(Fig.\ 2) the open box accelerator, after a period of acceleration, $t_a$, is 
transformed to what we call the ``convex floor amplifier'' by warping the floor, 
by pressure from underneath, to a 10 cm (x) by 50 cm (y) arc of a y-axis cylinder.  
The radius of the convexity would be several hundred m, projecting a few microns 
above the original plane in the middle.  In each of successive bounces on such a 
floor the two spin states, initially slightly separated in x by the accelerator's 
electric gradient, will bounce on slightly different slopes and move further apart 
in x, exponentially.  A computer calculation confirms that with a floor having a 
300 m radius, the initial $\Delta x$ will grow about a factor of 3.5 every second 
(a gain of 
500 in 5 s) independent of $v_z$, for $v_z >> v_x$, all the while maintaining coherence.  
The convex floor amplifier achieves a practical $\Delta x$ value in typically one to two 
orders less time than the accelerator alone.  The dihedral roof, with slopes of 
0.006 rad, will still retain for a useful amplifier time almost all UCN that were 
retained with a flat floor despite the box remaining open in x.  The convex floor 
amplifier's shorter cycle time means more runs, hence more UCN, hence better 
statistics in a given time.  
	
Table 1 exhibits a comparative example of each type of amplifier.  When the convex 
floor amplifier is used, its gain factor $G^{t_{CFA}}$ is inserted in the right hand side of 
Eq.\ (12); here is the gain per sec and $t_{CFA}$ is the time on the convex floor.  
We have calculated that a permanent convex floor will make the effective electric 
acceleration time very short because the gain quickly dominates and this 
time-shortening in turn greatly increases the relative importance of the 
undesirable ${\bf v} \times {\bf E}$ effect (see Sect.\ IV-A.)
	
To convert the amplified Æx to a phase difference, at an appropriate time 
the UCN are made to rise to the gravitational amplifier as described above.  
On the 45$^{\circ}$ mirror $\Delta x$ becomes $\Delta z$, thus generating a 
gravitational potential energy difference $mg\Delta z$ between the paired spin states.  
The time integral of this $\Delta V$, averaged over their now much shallower 
parabolic trajectories as they bounce on the floor of the shallow drift space box, 
generates a phase difference.  
	
The choice of having UCN escaping from the tilted floor upward rather than downward 
is made to simplify the floor-warping bar and its support.  The phase difference 
growth begins as the UCN leave the 45$^{\circ}$ mirror, and terminates at $td$ as 
the neutrons enter the Mach-Zehnder interferometer.  As mentioned in the introduction,
the beam splitter of the Mach-Zehnder must be a magnitized mirror which reflects
one spin state and transmits the other.  They are reunited at the non-magnitized
recombiner mirror through the interferometer, between the splitter and the recombiner,
one spin state must be rotated $\pi$ radians.  The phase difference between the two
states is given by the ratio of the counts in the two counters
of the Mach-Zehnder in the usual fashion.  One measurement
should be made with the electric fields in one direction and another with the fields in
the opposite direction.

\subsection{The Assembly}
We will describe the assembly which uses only the simpler gravitational amplifier.  
The general management of the UCN would be similar in the case of the convex floor 
amplifier.  The neutrons coming from the source must be x-polarized.  They could be 
led into the open box in the y-direction through a y-mirror gate.  It would take about 
1s to fill the box.  After this the RF-field needed to rotate the spin $\pi$/2 to achieve 
differential acceleration would be started and run for an appropriate time (a few seconds).  
	
After about 100 seconds in the open box, a 10 cm length (in y) of the floor would 
be tilted clockwise 0.01 rad.  The UCN would rise to the gravitational amplifier 
through a ceiling slot, hitting the 45$^{\circ}$ mirror (Fig.\ 2).  This takes 
advantage of motion in the y direction to improve transfer efficiency.  
The operation would take the order of 3 seconds in this case and result 
in a loss of perhaps 1/3 of the neutrons.  Reflection on (yz) planes during 
the transfer must be avoided.
	
The environment of this experiment must be a guide field of the order of 10$^{-8}$ $T$, 
as uniform as practical.  It must be in a vacuum chamber and it must be shielded 
as well as practical from external magnetic fields.  Some details of these 
requirements are discussed in Sect.\ IV.
	
After the second spin flip the UCN are less vulnerable to minor magnetic field 
irregularities and it would be best to lead them outside the magnetic shield and 
place the polarimeter there.  The point of this is that the polarimeter may need 
a rather high magnetic field and this could be disruptive inside the shield.
	
We can now make an estimate of the sensitivity of the proposed instrument, using $t_a$, 
$t_d$ and ($\partial E_x/\partial x$) from the gravitational amplifier column of 
Table 1 in Eq.\ (12).  In the table we arbitrarily assumed $\mu _e$ = 10$^{-28}$ e.cm, 
for which we get $\Delta P_e$ = 0.023.  Could one achieve that accuracy?  This of course 
depends on the number of neutrons that can be counted (and systematic errors, some 
of which will be discussed later).  The number of neutrons which could be counted 
in one filling should be given by

\begin{equation}
	N_c = V_b \,\rho \,DL\,v_s 
\end{equation}
where $V_b$ is the volume of the box, 2500 cm$^3$.  $\rho$ is taken as the phase space 
density of 
UCN attainable at the Institut Laue-Langevin (ILL) [1], which is 0.0036 cm$^{-3}$ 
(m/s)$^{-3}$.  
$D$ is the decay factor, 0.54, for 200 s [13], primarily due to inelastic collisions 
on the Fomblin surface coating (see Sect.\ IV-C) and $L$ is the fraction surviving 
after miscellaneous losses (transfer, counter efficiency etc.), guessed as 0.2.  
$v_s$, the velocity space occupied, requires some discussion.  It is intended to
be an average over the real space volume occupied in the open box.  The limiting 
velocity in x is 0.3 m/s.  We believe this should be reduced by a factor of 2 
for the average over the box.  In the y direction 0-4.5 m/s is usable.  A similar 
analysis of z suggests the limits 3 to 4.5 m/s.  All these ranges must be doubled 
for their negative velocity counterparts.  $v_s$ then becomes about 8.1 (m/s)$^3$.

>From all this we get $N_c$ = 7.9 counts per cycle.  While the rate is low 
it should be noted that the counts are concentrated in the last few seconds of 
a cycle, the live time of the counter, and should be well above background. 
We should also explain that we have chosen to use cycle times considerably less 
($\sim$ 0.2) than optimal for our case to get the counting rate up to practical levels.

The estimated fractional error in the polarization measurement would be
$2/(\alpha \sqrt{N})$,   
where $N$ is the total number of counts, half in each electric field direction.  
$\alpha$ is a ``polarization efficiency'' or ``fringe visibility'' factor 
(= 0.64 in Ref.\ 1).  
In Table I we have calculated the N which would give a statistical error (1s) 
equal to $\Delta P_e$.  Note that this value of N is the same as given by the 
energy-time uncertainty limit, 
$\Delta \epsilon \Delta t \geq 2\hbar /(\alpha \sqrt{N})$,
where $\Delta \epsilon$ is the relevant energy 
difference, 
i.e., the potential energy difference $mg\Delta z$, in the drift space.  
Actually realizing such a sensitivity is extremely difficult owing to the 
possible systematic effects, some of which are discussed in Sect.\ IV.

\section{Comparison to the Precession Method}
This comparison is clearly arbitrary.  We use for the accelerator method the 
parameters of Column 1 of Table 1 except in the case of the ratio of electric 
forces, $E_p/(\partial Ex/\partial x)$, which we conservatively estimate as 
1.0 at most on the basis of arguments concerning leakage currents in Sect.\ IV-D.  
For the precession method we use the parameters of the new ILL experiment 
[14] that may well improve their sensitivity by an order of magnitude.
	
For the precession method [1] the minimum measurable EDM given by the statistical 
limit is 

\begin{equation}
\mu _{ep} =  {\hbar \over 2\alpha _e E_p t_p \sqrt{N_p}} \,.
\end{equation}
The corresponding quantity for our method (using the gravitational amplifier) 
obtained from Eq.\ (12) and our estimate of the accuracy of our phase difference 
measurement, $2/(\alpha \sqrt{N_A})$ is

\begin{equation}
\mu _{eA} = {2\hbar \over 1.3 \alpha e (\partial E_x / \partial x) gt ^2_{\alpha}
t^2_d \sqrt{N_A}}
\end{equation}
We assume both methods use the same UCN source and the same overall operating time.  
The ratios (Precession/Accelerator) of volumes of the two experiments $R_v$, 
is 60/2.5 = 24 [14].  The ratio of retained-velocity spaces, $R_{vs}$, is estimated 
as 781/8.1 = 96.  This assumes deuterated polystyrene coatings for the 
precession experiment and Fomblin for us (Sect.\ IV-C).  The ratio of cycle times, 
$R_c$, is 100/200.  Finally, the ratio of surviving fractions of UCN, $R_{sf}$,
we estimate as 3.5.  (We added a factor of 2 in favor of the precession method 
for our losses in an extra transfer operation.)  Given these assumptions, 
$N_A/N_P = R_c/(R_vR_{vs}R_{sf} ) = 6.2 \times 10^{-5}$.  From Eqs.\ (19) and (20)

\begin{equation}
{\mu _{eP} \over \mu _{eA}} = {0.33 gt^2_{\alpha} t_d (\partial E_x / \partial x)
\sqrt{N_A} \over t_p E_p \sqrt {N_P}} \,.
\end{equation}
Using $t_d = t_a=t_P = 100$ seconds, we then get

\begin{equation}
{\mu _{eP} \over \mu _{eA}} = 240 \,.
\end{equation}
i.e.\ the accelerator method is more sensitive.

\section{Problems}
It would be inappropriate at this time to attempt to discuss in detail the 
many technological problems that may arise.   However we will discuss a few 
that we believe to be the most troubling  sources of systematic error or 
serious background.  The omissions are largely in areas of technology which
are common to many experiments, e.g., vacua, vibration control, temperature 
control, magnetometry and high voltage. 

\subsection{The {\bf v $\times$ E/c$^2$} Effect}
	
The relativistically produced magnetic flux density experienced by neutrons 
moving relative to an electric field is given by 
$B = {\bf v} \times {\bf E}/c_2$ 
(SI units).  
Its axially directed gradient $\partial B_x/\partial x = [(v_z\partial E_y/\partial 
x)-(v_y\partial E_z/\partial x)]/c^2$ 
generates unwanted differential forces in the x-direction on the bipolarized 
magnetic dipole moment (MDM) in the accelerator.  A major concern in these EDM 
measurements, as also in the precession method, is the fact that one cannot 
separate the relatively large ${\bf v} \times {\bf E}$ force 
$(\mu _m \partial B_x/\partial x)$ from that from the EDM 
$(\mu _e \partial Ex/\partial x)$ by turning off the electric field.  The 
${\bf v} \times {\bf E}$ effect must be reduced by careful cancellation.  We give 
here a detailed calculation suggesting that the ${\bf v} \times {\bf E}$ effect would not 
limit measurements of an EDM greater than about 2 $\times$ 10$^{-30}$ e.cm.
	
By the symmetry of our geometry (Fig.\ 2), $E_y$ is small and can be made more so by 
guard electrodes so the first term of $\partial B_z/\partial x$ becomes negligible 
(the $v_z$ bouncing 
also helps by making $<v_z > \rightarrow 0$.  $E_z$ is a more serious problem.  
The measure of its importance relative to the EDM force is the effective 
$(\partial E_z/\partial x)/(\partial E_x/\partial x)$.  This ratio of the gradients felt 
by the magnetic and 
electric moments has been calculated for the electrode configuration of Fig.\ 2 
and is shown in Fig.\ 5 for an x,z section through the open box.  
Using this in a computer program we estimate that the average of the absolute value of 
that ratio over the volume traversed by the retained UCN would be about 10$^{-3}$.  
This is a small number, but there is a worry here that this calculated number 
may be spoiled by nonuniformities in the dielectric constants of the top and 
bottom plates of the stationary box.  Making these plates much wider than 10 
cm as shown in Fig.\ 2 should reduce edge effects.  (Note that the dihedral 
ceiling angle would not be extended by more than 5 cm from the midplane however 
so that UCN which entered the high $\partial E_z/\partial x$ regions would not be 
retained.)  
Simple measurements of the dielectric constant of the plates could be made to get some 
idea of their uniformity.  In addition it should be possible, although perhaps 
difficult, to measure $\partial E_z/\partial x$ directly using a piezoelectric crystal.  
This would furnish a second check on the plates.  This problem of uniformity 
would probably determine the material used for the plates.
	
The ${\bf v} \times {\bf E}$ effect is only important in the accelerator.  After that 
the amplifier quickly raises the energy difference so much that ${\bf v} \times {\bf E}$ 
does 
not matter.  The trajectories of the UCN in the box will naturally lead to 
considerable cancellation of the second term in $\partial B_x/\partial x$, due to scanning 
of the antisymmetrical $\partial E_z/\partial x$.  Using the program for calculating 
trajectories and $\partial E_z/\partial x$ we have calculated the $\Delta x_{\alpha}$ 
induced 
by the ${\bf v} \times {\bf E}$ effect for a wide range of initial conditions that still 
leave the UCN confined in  the 10 cm width of the open box for 100 seconds.  
The resulting $\Delta x_{\alpha}$ varied over a 100:1 range, the largest tending to 
be those 
passing through high $(\partial E_z/\partial x)$ regions.  The average of the 
largest 1/4 of the $\Delta x_{\alpha}$ from the ${\bf v} \times {\bf E}$ 
effect produced approximately the same
$\Delta x$ as an EDM of $\sim 10^{-28}$ e.cm.  Since that $\Delta x$ is 
of random sign the 18,500 
neutrons, we claim could be counted in 5.4 days, allow a measurement of the EDM to 
$\sim 2 \times 10^{-30}$ e.cm accuracy, as far as the ${\bf v} \times {\bf E}$ effect 
is concerned.  
These results are considerably lower than a random walk approximation in which 
we took 5 cm steps and in which the force was random but the velocity was allowed 
to accumulate.  In our actual case the forces are not random because they tend to 
change sign systematically about every two cm.  This shows that the systematic 
cancellations are having a very appreciable effect.  The ${\bf v} \times {\bf E}$ 
effect on $\Delta x$ 
appears to rise linearly with time where the EDM of course goes as $t^2_{\alpha}$.
	
It is fairly clear from looking at Fig.\ 5 that the relative size of the 
${\bf v} \times {\bf E}$ 
effect versus that of the EDM could be reduced by a factor of 10 by reducing the 
width of the open box from 10 to 5 cm.  This would cost a factor of two in volume 
and two in range of $v_x$.  This in turn would only cause a factor of two loss in 
sensitivity if background were not a serious problem.
	
It should be said that the computer programs on the results of which these 
conclusions are based do not include the dielectric effects of the materials 
of the ceiling and floor of the open box.
	
There are several experimental approaches to measuring the ${\bf v} \times {\bf E}$ 
effect.  
One way is to measure the effect on $\Delta \phi$ of reducing $v_y$ by a factor of two.  
Another is to move the electrodes in by several cm.  The percentage increment 
in the ${\bf v} \times {\bf E}$ effect would be roughly 2.5 times that of the EDM.  
One could also change $t_a$ by a factor of 1.4.  The EDM effect would change 
by a factor of two but the ${\bf v} \times {\bf E}$ by 1.4.

\subsection{Magnetic Fields}
(We will discuss magnetic fields due to leakage currents in Sect.\ IV-D.)  

\underline{Components}.  The apparatus will need very good magnetic shielding, 
perhaps even superconducting shields.  Our estimates based on the experience of 
the EDM group at ILL indicate that conventional magnetic shielding should enable 
interferometric detection at the 10$^{-26}$ e.cm level, and that the use of 
superconducting shields might allow reaching substantially lower.  Their 
analyses showed the largest expected systematic effect (spurious EDM) arises 
from hysteresis in the conventional high permeability shield caused by current 
pulses as the electric field is reversed.  Such small displacement currents would 
not affect superconducting shielding.  To limit a spurious EDM signal to 10$^{-28}$ e.cm 
their precession method needed a magnetic field difference between the runs with 
reversed electric field averaged over the months' long duration of the experiment 
of 0.5 = 10$^{-17}$ $T$.  In their experiment they got a shielding factor (SF) of 
10$^5$ which, for example, reduced the field from a 20 cm radius current loop carrying
20 mA at a 2 m distance from their shield to the tolerable level, 10$^{-11}$ $T$ 
at the shield outer surface.  A simple one-layer superconducting shield should give 
SF $>$ 10$^8$ even for much larger ambient disturbances.  A 2m internal 
diameter superconducting shield will surely be the most expensive part of the 
apparatus.  A brief discussion of a scheme that may improve superconducting 
shields appears in Ref.\ [16].
	
A guide field of the order of 10$^{-8}$ $T$ will be needed to keep the spins of the 
UCN oriented.  This leads to a Larmor precession of 1.8 radians per second.  
(ILL uses 180 radians per second.)  In principle the phase shift this induces 
could be ignored since it is nearly the same for all UCN and could be removed 
by the second $\pi$/2 spin rotation in an appropriate fixed phase relation with the 
first (Sect.\ II-A).
	
\underline{Perturbing magnetic fields}.  In our proposed EDM measurements all 
sources of non-uniformities in the magnetic field, particularly in the accelerator, 
lead to unwanted phase differences.  They are of serious concern because of the 
huge ratio of $\mu _m / \mu _e$.  We believe it is only the x-component of the 
field which 
contributes significantly because the UCN are polarized in that direction.
	
There are two such phase-difference-generating terms of concern.  
The first gives an increment in the precession rate, hence in $\Delta \phi _p$,

\begin{equation}
\Delta \phi _p = {1\over \hbar} \int ^{t_a + t_d}_0 2\mu _m \Big( B_x -
\langle B_x \rangle _s \Big) dt \,.
\end{equation}

Here $<B_x>_s$ is the spatial average of $B_x$ in the accelerator and drift spaces.  
The second and more serious disturbance is a magnetic gradient term significant 
only in the accelerator because it produces a $\Delta x_{ag}$.  The effect of this term 
is amplified in the gravitational amplifier as is the $\Delta x_a$ from the EDM.   
In analogy to Eq.\ (11),

\begin{equation}
\Delta \phi _g = 0.67 {g\mu _m t_a \over \hbar} \int ^{t_a}_a \int ^{t_d}_{t_a}
{\partial B_x \over \partial x} (dt)^2 \,.
\end{equation}

Since numerically the gradients and the field non-uniformities are of the same order, 
the $gt_d$ factor, of the order of 1000, is very dominant, and the precession effects can 
be ignored.  The residual x-directed magnetic gradients may well impose the most 
severe limit on the attainable EDM sensitivity, beyond the constraints of the 
${\bf v} \times {\bf E}$ effect, even after considering that the phase shifts from magnetic 
gradients are more easily distinguished from those produced by the EDM, by turning 
off the electric field.
	
The different paths taken by the individual UCN will lead to different values of 
$\Delta \phi _g$ because the gradients are not uniform.  This could be a serious 
problem because 
polarization, which is what would be measured, is a sinusoidal function of 
$\Delta \phi _g$ (Fig.\ 6).  
The $\Delta \phi _g$ of the ensemble must be confined to about $\pm$1 rad and centered near 
$P_x$ = 0 on one side of a fringe where $P_x$ is a reasonably linear function of 
$\Delta \phi _g$.  
If the standard  deviation of the $\Delta \phi _g$ in an ensemble were as much as 5 
rad the 
$P_x$ would be that from an average over several fringes and $|P_x (max)|$ would be 
very small and hence the slope, which determines the sensitivity to the EDM 
would also be very small.  This would then require an enormous number of UCN 
to measure the much smaller effect of the EDM with even modest accuracy.  
If the average value of $\Delta \phi _g$ is appreciable, it may be necessary to adjust the 
phase of the second spin flip slightly to keep the phase difference centered near zero.
	
It should be noted that Eq.\ (24) is probably a random walk since there are likely 
to be many sign changes of $\partial B_x/\partial x$ scanned in the course of 100 s.  
It would be a sort of integrated random walk, much as we had with the 
${\bf v} \times {\bf E}$ 
effect and $\Delta \phi _g$ would grow as $t_a$, not as the EDM effect which grows as  
$t^2_a$.
	
There are three possible sources of magnetic gradients; ferromagnetic 
impurities in the apparatus, external fields penetrating the shield and 
non-uniformities in the guide field itself.  Leakage fields must be limited 
but no usable shield can be totally closed, hence none is perfect.  It may be 
desirable to have several monitors of the magnetic field outside the shield and 
when these external fields are beyond some limits, shut the measurements down.

We have done some experimental model studies on the problem of field uniformity 
of the guide field [16] using a Helmholtz-derived  double four-coil, 
guide-field-generating assembly fitting inside a simulated superconducting shield.  
The measured field inside the accelerator box was uniform, i.e.\ its variation was 
not detectable at a sensitivity of 10$^{-4}$.  $B_x$ calculations on the double four-coil 
assembly gave an average gradient in the accelerator of 10$^{-6}$ (x10$^{-8}$) T/m and 
the scanning for 100 s by the UCN will, we estimate, reduce the magnetic 
$\Delta x$ production by a factor of at least 40, giving an effective gradient of 
2.5 $\cdot$ 10$^{-16}$ T/m and of random sign for each UCN.  18,500 UCN per batch (Table 1) 
will reduce the variation of the effective magnetic gradient to 0.7\% of 
2.5 $\cdot$ 10-16 $\rightarrow$ 2 $\cdot$ 10-18 T/m, hence producing $\Delta x_a$ 
(and $\Delta P_x$) equal to that from an EDM of 3 $\cdot$ 10$^{-29}$ e.cm.  
This can be checked by measurements made with UCN using the techniques 
described below.  (The electric fields must be off, of course.)
	
\underline{Reduction of Effects of Variations}.  
The variation of the magnetic fields can be 
examined with SQUID gradiometers [17] but the ultimate test will have to be done with 
UCN.  This could be done by measuring $P_x$ as a function of frequency (Fig.\ 6).  
If, in measurements at, say, four frequencies giving points 1 radian apart in 
phase at the second $\pi$/2 flip, one gets $P_x$ near zero at each point this would be 
a sign of a large range of values of the gradient integrals.  Reduction of 
sensitivity by reducing the
time, $t_d$, by a factor of $\sim$5 should help to increase $|P_x(max)|$.  
But one must eventually remedy the magnetic non-uniformities if one needs the 
full potential sensitivity to the EDM.  
	
In general there are two approaches to coping with excessive magnetic gradients; 
diagnosing their origin and reducing them, or, lowering the sensitivity of the 
apparatus to them relative to the EDM sensitivity without unacceptable loss of 
the latter.  Reversing the guide field will show its influence by use of a SQUID 
gradiometer.  
One could discriminate between shield leakage and ferromagnetic impurities by 
varying the external field.
	
Changing the 4-coil guide-field assembly [16] to an 8 coil assembly would give 
a guide field uniformity at least tenfold better, in principle.  Whether it would 
be that good in practical construction is an open question.  Gradients this small, 
10$^{-14}$ T/m or a field difference of $5\cdot 10^{-16}$ T over x = 10 cm of 
the open box, 
on a $<B>$ field of 10$^{-8}$ T can be measured by a SQUID gradiometer.  The residual 
gradient effect will have to be assayed by measuring its effect on 
$\Delta  P_x$ by UCN polarimetry.
	
Guide field gradient correction coils in the shield in series with the guide field 
coils are a possibility for all the guide field gradients of importance.  Serious 
gradients from ferromagnetic inclusions could possibly be handled similarly with 
independent current sources.  The design and test principles are discussed for a 
simpler case in Ref.\ [18].   A strong effort should be made to achieve mirror 
symmetry about the x-midplane in relevant structures.
	
Increase of the guide field current when one is trying to measure the gradients would
probably be helpful.  Such increases are limited of course by the danger of thermal 
expansion owing to heating by the current.
	
Reduction of the x-length of the accelerator by a factor of two would very 
likely give a major reduction in the gradients.  A serious reduction of the 
guide field current would reduce the gradients from the guide field of course 
but such a reduction is limited by the Larmor period increases.  The time spent 
in two spin flips, at least 6 Larmor periods, should be small compared to $(t_a + t_d)$.
	
Generating sharp, non-adiabatic reversals of the guide field at symmetrically 
distributed times in each run might be of use.  Even one reversal in the middle of 
a run of $t_a + t_d$ is helpful.  (This is equivalent to the ``$\pi$'' laser pulse in 
Ref.\ [7].) 
Reversals also make the phase difference less sensitive to the absolute value of the 
field and so more reproducible from run to run.  We believe that reversal, with as 
high a frequency as depolarization will permit, might reduce the variation in the phase 
difference generated by the guide field gradients.
	
Another possible tactic is to increase the length of time in the accelerator, 
perhaps to $t_a$ = 500s to gain x25 and decrease $t_d$ to 4 s, hence reduce the gain 
in the amplifier by 25.  This would give the EDM the same overall gain but a 
gain increase relative to that of the magnetic gradients of a factor of 5 because 
the EDM effect grows as $t_a^2$ and the magnetic gradient effect grows more like $t_a$.  
The penalty is a factor of 2.5 loss in counting rate.  
	
Time variations of the fields must be held to 1 part in 10$^5$ to avoid smearing out the EDM 
effect.  This is well within the state-of-the-art of current control, but is not trivial.  
The structure of the coil should be quite stable as it is in a volume of well controlled 
temperatures and the possible small vibrations should not be dangerous because their 
effects average out.  In general, time variations are much less important than they are 
in the Ramsey method because our very low Larmor frequency, and hence small number of 
precession periods, makes our $\Delta \phi$ much less sensitive to magnetic field changes.

\subsection{Reflectivity}
	Fomblin is a fluorinated dehydrogenated nonvolatile liquid ether which has been shown 
[19] to reflect UCN with velocities below 4.55 m/s with exceptionally low inelastic 
collision loss rates ($<10^{-5}$) enabling UCN trapping lifetimes of the order of 3000 
seconds [15].  It is the material of choice for coating all the UCN guide channel 
and mirror surfaces requiring total specular reflection as it provides an adherent 
smooth film.  It seems likely that it will also improve the specularity of reflections 
as a coating on less-than-ideally planar or smooth surfaces by filling in voids, thus 
contributing significantly to cost reduction in producing optical flats.  
	
Fomblin does raise a possible difficulty because it would form meniscuses at 
corners.  The curved surfaces of the meniscuses could mix components of velocities
 such as $vx$ and $vz$.  There are several possible ways of mitigating this.  There may 
be a film thickness great enough for good reflection  but too thin to form a 
meniscus particularly if there were a several micron separation of the orthogonal 
surfaces at the corner.  A more heroic possibility is the use of frozen Fomblin or 
frozen O$_2$ surfaces as in a remarkable neutron-lifetime measurement done in Russia 
[20].  This appears to have worked very well. O$_2$ gives a lifetime for UCN comparable 
to Fomblin's and a neutron velocity cutoff substantially higher than Fomblin's.

\subsection{Leakage Currents and Surface Charges}
Leakage currents can lead to magnetic field gradients which can accelerate the UCN 
through their relatively huge MDM.  Their effects must be minimized, as in the standard 
precession EDM measurements, which sets effective limits to the field strengths that 
can be used.  These currents can be measured in sum at the source; tracing their branching 
paths is more difficult.  In general we believe this problem is less serious here than in 
the precession method.  In that method the full voltage is applied across a 20 cm section 
of insulation which must also function as a bottle wall.  In our case the insulators 
could be $>$50 cm long and are less constrained by non-electrical requirements. 
Moreover the leakage paths are relatively remote from the UCN trajectories.
	
Related effects can arise from the fields and currents due to surface charges 
and their motion relative to the UCN.  These charges are probably minimized by 
coating the surfaces with the liquid Fomblin which would make them mobile and 
prevent their buildup.  However, it is difficult to be quantitative about this 
and experimental study is needed.  These effects are similar to the 
${\bf v} \times {\bf E}$ 
effect (Sect.\ IV-A) in that it is difficult to separate them from the 
EDM effect because they both depend on the electric field.  They can perhaps be 
separated by their possible non-linear dependence on field strength.

\subsection{Precision and Stabilization}
	The other components of the interferometers are standard parts of other UCN experiments 
such as the precession EDM measurements; polarizing mirrors and foils, plane mirrors, 
magnetized foils, an RF spin flipper, and detectors, although the demands on precision 
and stability will surely be more severe.
	
The specifications of flatness, angular tilt and stability should be less demanding 
than for light interferometry because the partial waves almost completely overlap.  
There is some concern that reflections which may slightly interchange velocity 
components should not significantly affect $\Delta v_x$ or $\Delta x$.  However, 
the coherent bipolarized state essentially ensures this.
 
\subsection{Gravitational and Sagnac Effects}
Avoiding unwanted mixing of $v_x$ and $v_z$ in the UCN trajectories in the 
gravitational field is simplest if the floors of the accelerator and drift 
spaces are horizontal.  A Fomblin pool floor would help here.  As the tiny 
separation of the partial wave packets is along x in the accelerator, the z 
separation, hence the gravitational and Sagnac (Coriolis) relative phase shifts 
are essentially zero at all times.  In the drift space the $\Delta z$ intentionally created 
at the 45$^{\circ}$ mirror of the gravitational amplifier will oscillate 
between variable $\pm$ values with each bounce of the partial wave trajectories 
on the Fomblin floor as they cross vertically.  While $<\Delta z>$ is not zero, such Sagnac 
contributions to the total phase shift produced in the drift space are negligible 
(10$^{-4}$) compared to the gravitational acceleration that is amplifying the EDM-generated 
shift.

\subsection{Decoherence}
The possible loss of coherence of the two spin states of a given UCN is of 
some concern.  This would presumably occur at a reflection as it is hard to see what 
might cause it in a vacuum.  It would probably take an inelastic scattering to cause 
decoherence but almost all such result in an increase of the UCN energy since the UCN 
are far colder than the apparatus (I mK vs.\ at least a few K or, more likely, room 
temperature).  Above several mK the neutrons do not reflect well and are quickly lost.  
The ILL experiment [1] is somewhat reassuring here.  They obtained clearly defined 
fringes (polarization vs,\ frequency of the spin-flip field) after 68 seconds in their 
chamber.  This also suggests no other source of significant random variation in 
phase differences exists at their sensitivity level.

\subsection{Control, Calibration and Alignment Measurements}
	
Systematic errors are probably the major worry in an experiment which purports to 
be as sensitive as the one proposed here.  We obviously cannot anticipate all of 
them, and perhaps not even the most serious ones in the absence of actual experiments, 
but a few remarks beyond what we have said about the ${\bf v} \times {\bf E}$ 
and magnetic field 
effects seem obligatory.  If a non-zero effect is found it would be advisable to 
show that it is independent of the sign of the electric field gradient and linearly 
dependent on the electric field and the drift time.  It should also be quadratically 
dependent on the time in the accelerator.  Some of the variants we have suggested 
in this paper have possible value as control experiments.  For example one could 
check the effect of varying the sign of the initial polarization or the strength of 
the guide field or the guide field reversal frequency etc.  If a non-zero EDM
$>$10$^{-28}$ e.cm is found, the potentially great sensitivity of the interferometer 
can be used to check for certain systematic effects.  Leakage currents could be 
checked by reducing the voltage and ${\bf v} \times {\bf E}$ by reducing $t_a$ 
to which the ${\bf v} \times {\bf E}$ and EDM responses are different.
	
One must also worry, if these extreme levels of sensitivity are reached, if one is not 
confusing a new physical effect with a systematic error.

An internal monitor on the overall instrument performance is the magnetic 
moment of the neutron.  Coils placed near the accelerator cell can generate 
known weak magnetic field gradients with spatial configurations similar to 
the electric gradients.  Bipolarized UCN can thus be accelerated by their magnetic 
moments and put through the entire instrument.  A comparison made between the 
measured and predicted phase shifts will serve to reveal a number of systematic 
effects distorting the EDM measurements but not those arising from the electric 
field, i.e.\ current leakage and ${\bf v} \times {\bf E}$ effects.  
Both electric and magnetic
acceleration can then be concurrently opposed by varying the coil current and 
its polarity to seek a null phase shift.  This comparison method will calibrate 
the sensitivity, give the sign of the EDM unambiguously, compensate for any 
non-linear phase shifts, and allow the maximum sensitivity to be used to give the 
highest EDM measurement precision.  

\section{Simplified Test of the Operating Principles}
	
We have noted that preliminary tests of the novel principles of our proposal and 
of their combination are desirable.  The relatively large magnetic moment of the 
neutron makes possible two such simplified and relatively inexpensive tests, each 
being a rough measurement of the MDM, using differential acceleration of bipolarized 
neutrons by a magnetic gradient, and drifting to develop a phase shift.  The first, 
and simplest, tests these principles for (thermal) neutrons (which the Kasevich and 
Chu experiment [7] has established for sodium atoms).  As noted in Sect.\ II-C, 
there has been criticism of our claim $\Delta \phi$ grows during our drifting stage.
	
\underline{Linear Drift Space Test (Fig.\ 7)}.  Horizontally polarized neutrons, 
v = 2,000 m/s, selected either by monochromatization or by time-of-flight 
measurement of a pulsed source, flow along x, the axis of a long solenoid 
equipped with y axis $\pi$/2 RF-flipper coils near each end.  The solenoid generates 
a guide field, $Bo$ = 10$^{-3}$ $T$.  The current in the central region, I-0, is held 
constant to 0.1\%.  The currents in the two end regions within the flip coils, $A$ 
and $B$, can be (equally) varied, to give $\Delta B$ = 10$^{-5}$ T, to produce magnetic 
gradients 
at I and 0.  The A flip coil produces the bipolarized state whose MDM is 
differentially accelerated at I, drifts in the 0.5 m drift space I-0, is decelerated 
at 0 and is terminated in flip coil $B$, and thence its polarization is measured.  
The experiment consists of measuring the polarization shift which is the difference 
$\Delta P$ with $\Delta B$ on, minus $\Delta B$ off.  The initial polarization 
is set to zero 
($\pm$0.05) by adjusting the solenoid current with $\Delta B$ off.  With a similar setup, 
Robert {\it et\,al}.\ [21] have shown interference with metastable hydrogen atoms.

With $\Delta B$ on, the polarization changes, due to two effects.  The first is the 
effect of 
the acceleration at I and the motion through the drift space, I to O.  This should 
give a phase difference calculable from Eq.\ (3), and a polarization change of sin 
$\Delta \phi$.  For the parameters we give this yields a polarization change of about 0.45.  
In addition, there is a smaller polarization change due to the change in precession 
in the regions A to I and O to $B$ when $\Delta B$ is turned on.  This is easy to 
calculate to sufficient accuracy and for the case described produces a polarization 
change of
 
\begin{equation}
\Bigg( {\overline{AI} + \overline{OB} \over \overline{IO}} \times 
{\Delta B\over B_0} \times \int^B_A 2 \mu _m B_x dt \Bigg) \approx 0.1 \,.
\end{equation} 
Here $\overline{AI}$ means length from $A$ to $I$.  This correction can be applied 
to the measured $\Delta \phi$ to yield the effect of the phase generation due to drifting.
	
\underline{Miniaturized UCN Test}.  
The second experiment using UCN would address the same questions 
plus the retention of UCN in an open box with a dihedral roof and the functioning of a 
gravitational amplifier and drift space with two active coordinates (x and z).  
The strong magnetic acceleration of the magnetic moment makes possible a UCN test 
on a greatly reduced hence cheap (cost $\sim$ a few percent of the real experiment) scale.
	
The UCN would be retained briefly ($\sim$1 sec) in a small open box with dihedral ceiling 
with magnetic gradient coils followed by a gravitational amplifier and $\sim$1 sec. drift, 
then $\pi$/2 flipping and polarization detection, all in a short cycle $\sim$2 sec.\ 
repeated frequently.  This could be done in a small ($\sim$20 cm) apparatus with 
simple magnetic shielding and possibly in a 99
enclosure.  One hours' data accumulation should yield a sufficient test of the 
agreement of measured and known MDM.  A pulsed (chopped) UCN input beam with the 
applied magnetic gradient turned off (or reversed) on alternate pulses (which are 
separately recorded) would promote a good statistical subtraction of the phase 
shift arising from a, possibly varying, ambient magnetic gradient.
	
It is possible that a considerably cheaper test could be conducted using polarized 
$^3 He$ [22].  Because the phase space density in $^3He$ could be enormously greater 
than that of UCNs, the apparatus could be much smaller and more easily shielded and 
still have higher counting rates.  It also avoids the numerous problems of working at 
a reactor.

\section{Other Interferometers}
	
There are several other types of interferometers we have considered.  As a warning 
we will describe a superficially attractive method which does not work.  
One would introduce a broad spectrum of bipolarized UCN in a toroidal chamber, 
moving circumferentially (Fig.\ 8).  A concentric, roughly toroidal magnetic guide 
field keeps the two spins approximately circumferential.  The two UCN spin states 
would be accelerated in opposite circumferential directions as they passed through 
the electric field gradients at the top and bottom of the figure and nominally could 
do this indefinitely.  Unfortunately the radial gradient of the magnetic field - inherent 
in its curvature-supplies exactly the opposite acceleration.  How can this balance an 
independent electric field effect?  It does so because the relevant circumferential 
magnetic acceleration is due to a surprising radial component of the magnetic moment, 
which component is induced by the electric field acting on the EDM.

\section{Conclusions}
	
Let us admit that this entire paper is very optimistic as to both the technical and 
economic limits of an interferometer measurement.  However, there is an additional 
argument for the interferometer that should be mentioned.  Systematic errors common 
to two experiments are usually less serious in the one where the signal has been 
increased even at the expense of some loss in statistical accuracy.  Specific problems 
of a given experiment can of course affect this generalization drastically.  
Systematic errors may well dominate future developments of the EDM measurements.
We believe this is a promising proposal but some important features are not firmly 
established.  What we are arguing here is that they are worth testing as in Sect.\ V.

\acknowledgments
	We are deeply indebted for the help we got from the late Leonard Goodman.  
It is a pleasure to be able to thank Henning Esbensen, Morton Hamermesh, 
Edward Hinds, Victor Krohn, Murray Peshkin, Norman Ramsey, Helmut Rauch, 
Jack Uretsky and Samuel Werner for advice and help.  
We have also benefitted from a paper by S. K. Lamoreaux and R. Golub \cite{lamoreaux}.
One of us (T.D.) 
wishes to acknowledge the generous support of Argonne National Laboratory as a 
Special Term Appointment in the Physics Division during the time of the completion 
of this manuscript.

\bigskip
\indent
This work was supported by the U.S. Department of Energy, Nuclear Physics Division, 
under Contract No. W-31-109-ENG-38.

\newpage
\section*{Appendix}
This appendix is intended to demonstrate that in the open box accelerator the 
difference in velocity of the bipolarized states does not change sign when the 
pair reverses its x-velocity.  In Fig.\ 9 we show the velocity vectors in successive 
reflections on the dihedral ceiling, the right hand one reversing their x-motion 
component.  Here $\Delta \bf{v} = \bf{v}_{\rightarrow} - \bf{v}_{\leftarrow}$;
$\bf{v}_{\rightarrow}$ 
is the velocity of the spin state that is 
always electrically accelerated to the right and $\bf{v}_{\leftarrow}$ 
the velocity of that always 
accelerated to the left.  The points of reflection of the wave vectors are minutely 
separated in reality but have been moved together for clarity.  The tiny effects of 
further electric acceleration play a completely insignificant role in the time span 
of the events described below.  The following applies for either direction of motion 
and for either ceiling slope or for any combination thereof.
	
The vector triangles $\bf{v}_{\rightarrow}$, $\bf{v}_{\leftarrow}$ and 
$\Delta \bf{v}$ for the incident and reflected waves are mirror 
images reflected in the plane normal to the ceiling.  $\rho$ is the angle at any time 
between the vertical and the incident $\bf{v}_{\rightarrow}$
(or $\bf{v}_{\leftarrow}$  since they are almost perfectly 
parallel).  $\bf{v}_{\rightarrow}$ and $\bf{v}_{\leftarrow}$  
are almost perpendicular to $\Delta \bf{v}$ hence $\rho$ 
is also the angle between $\Delta \bf{v}$ and 
the horizontal ($\rho '$ for the reflected  beam), $\theta$ 
is the dihedral slope angle, 0.006 rad.  
On reflections not reversing x-motion, e.g., the left one in the figure, it can be seen 
that the angles I, R and $\theta$ are related by  
$I = \rho - \theta = R = \rho ' + \theta$, so $\rho ' = \rho - 2\theta$ 
for waves 
moving away from the x-midplane of the open box (x=0), either to right or 
left, and $\rho '= \rho +2 \theta$ when moving toward x = 0.  
Hence after n ceiling reflections 
(n $\leq$ 6) when receding from x = 0, $\rho '_n = \rho _1 - 2n\theta$ 
will change sign, therefore the 
x component of motion will reverse as in the right hand reflection.  Before and after 
reflecting in x, 
$\bf{v}_{\rightarrow} = \bf{v}$$_{\leftarrow}$ + $\Delta \bf{v}$ 
which shows that $\Delta \bf{v}$ continues to point right even 
though the UCN will now move left, i.e., $\Delta \bf{v}$ does not reverse when the 
bipolarized 
UCN reflects its x motion.  As the x-reflection occurs the vectors
$\bf{v}_{\rightarrow}$ and $\bf{v}_{\leftarrow}$ are 
most nearly vertical so $\Delta \bf{v}$ is nearly horizontal, and as the UCN passes x=0 
the $\Delta \bf{v}$ vector tilts at most $2n\theta = 8^{\circ}$ from horizontal 
and then the tilt diminishes.  
At x=0 the electric gradient force is briefly reduced at most $(1-cos8^{\circ})$ = 1\%.  
The oscillation by $2n\theta$ continues until $t_a$.
	
Note that the center of mass of the bipolarized state of the neutron experiences 
no acceleration in any type of electric or magnetic accelerator with static fields.  
In both open and closed box accelerators, as each partial wave recrosses any 
electrical equipotential surface, it will have exactly the same kinetic energy, 
ignoring gravity, as on previous crossings, i.e., it will have been accelerated 
and equally decelerated between successive crossings.  In the open box 
it nevertheless can gain $\Delta v_x$ and $\Delta x$ with respect to the other partial wave 
(which will have been decelerated, then accelerated) because the two spin states 
do not cross such equipotential surfaces at the same instant, having had different 
$<v_x>$.

\begin{table}
\caption{Parameters of the Interferometers.}
\begin{tabular}{llll}
\multicolumn{2}{r}{Electrode Potential Difference} &
\multicolumn{2}{l}{100 kV} \\
\multicolumn{2}{r}{Electrode Spacing} &
\multicolumn{2}{l}{0.5 m } \\
\multicolumn{2}{r}{$\partial E_x/\partial x$} &	
\multicolumn{2}{l}{1.1 $\times$ 10$^6$ V/m$^2$}  \\
\hline
& Gravitational & Convex Floor & Units \\
& Amplifier & Amplifier & \\
Assumed EDM & 10$^{-28}$ & 10$^{-28}$ & e$\cdot$cm \\
Assumed $t_a$ & 100 & 15 & s \\
Assumed $t_{CFA}$ $^*$ & NA & 15 & s \\
Assumed $t_d$ & 100 & 5 & s \\
$\Delta x_a$ $^{**}$ & 1.05 & 0.024 & pm \\
$\Delta x_d$ $^{\dagger}$ & 0.26 & 0.31 & nm \\
$\Delta P_e$ & 0.023 & 0.027 & NA \\
$(\Sigma t )_c$ & 200 & 30 & s/cycle \\
$D$ & 0.54 & 0.91 & decay factor \\
$N_c$ & 7.9 & 13.3 & counts/cycle \\
$N$ & 18,500 & 13,900 & total counts \\
Duration & 5.4 & 0.36 & days \\
\hline
\multicolumn{4}{l}{$^*$Time in the convex floor amplifier.} \\
\multicolumn{4}{l}{$^{**}\Delta x$ after acceleration.} \\
\multicolumn{4}{l}{$^{\dagger}\Delta x$ after drifting.} \\
\end{tabular}
\end{table}

\begin{figure}
\caption{Flow diagram of a simplified (schematic) interferometer.  
The arrows denote the EDM vectors and their relative orientation and motion 
(exaggerated).  To convert to a more useful assembly (see Fig.\ 2), the dashed 
components would be inserted.}
\end{figure}

\begin{figure}
\caption{Vertical cross sections of accelerator and amplifiers.  The slopes of 
the dihedral ceiling (and of the floor in the transfer position) are exaggerated.  
The electrodes are shown at half the distance from the center of the figure they 
would have in the apparatus.  The UCN source, the polarizer and Mach-Zehnder
interferometer are not shown.}
\end{figure}

\begin{figure}
\caption{	Electric fields and gradients in the open box.  Units of fields are arbitrary.}
\end{figure}

\begin{figure}
\caption{a) The growth of the phase difference (of one of the ensemble of UCN pairs) 
generated to equal time in the drift space; b), the trajectories of the particular 
pair of bipolarized UCN which produced the phase difference.  Their minute separation 
is grossly exaggerated.  Averaged over the ensemble $\Delta \phi$ grows monotonically; 
c) the $\Delta z$ function (magnified) whose time integral generates phase difference.}
\end{figure}

\begin{figure}
\caption{($\partial E_z/\partial x$)/($\partial E_x/\partial x$) from our four 
electrode array, a measure of the ${\bf v} \times {\bf E}$ effect versus that of the EDM.  
The heavy and dotted lines outline part of the open box occupied by the UCN and 
the contours are isoratio lines.  Units are a ratio of 10$^{-3}$.  The shaded areas 
are all below $1\times 10^{-3}$.  Note the antisymmetry.}
\end{figure}

\begin{figure}
\caption{Idealized ``fringe'' plots.  $\sigma _g$ is the variance of the magentic
gradient integrals.
The $\Delta \phi$ is that after the second 
spin flip and for a magnetic field of 10$^{-8}$ T and a 200 s time interval between 
the first and second spin flip, the flips being each $\sim \pi$/2.  The RF frequency 
of the spin flip = 0.29164 Hz.}
\end{figure}

\begin{figure}
\caption{Thermal neutron test of differential accelerability of bipolarized 
neutrons and of phase shift development in the I-O drift space.  The 
accelerating magnetic gradients at I and O are developed by diverting 1.0\% 
of the constant guide field current from the short (0.1 m) ends of the 0.7 m 
solenoid.  The y axis $\pi$/2 RF flipper coils are centered at A and B, 0.6 m apart.  
Effects due to precession in $B_o$ can be measured by eliminating the field gradient.  
The RF spin rotation coils can operate continuously.}
\end{figure}

\begin{figure}
\caption{The toroidal accelerator.  The grounded electrodes carry the 
low-voltage current for the roughly toroidal guide field.  The current 
would be into the plane of the figure in the central electrode and out of 
the plane in the other two.  Two fragments of typical UCN trajectories are 
shown dotted.  The plane of the figure is horizontal.}
\end{figure}

\begin{figure}
\caption{Diagram in velocity space showing on the left a typical reflection 
on the ceiling and on the right a reflection where $v_x$ reverses.  Note that on 
the right the primed (reflected) velocities from the tilted ceiling represent 
leftward motion.  $\theta$ is shown increased by a factor $\sim$25; 
$\Delta \bf{v}$ and $\Delta \bf{v'}$ are 
increased by many orders of magnitude.  The dashed triangle shows what the 
reflection would have been from a horizontal ceiling.  Variables are defined 
in the text.  Note also the progressive leveling of $\Delta \bf{v}$ toward the right.}
\end{figure}

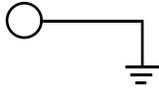
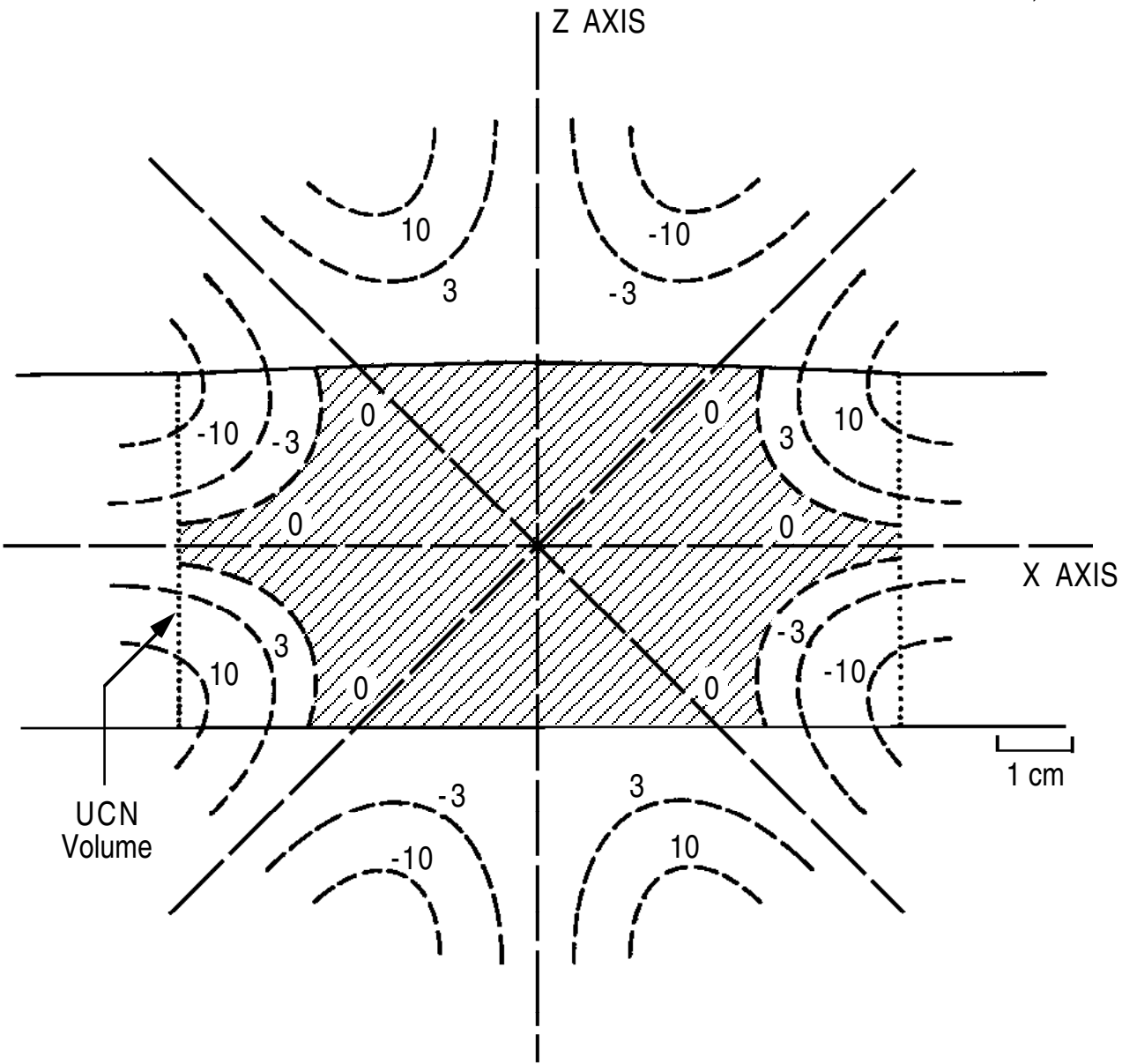
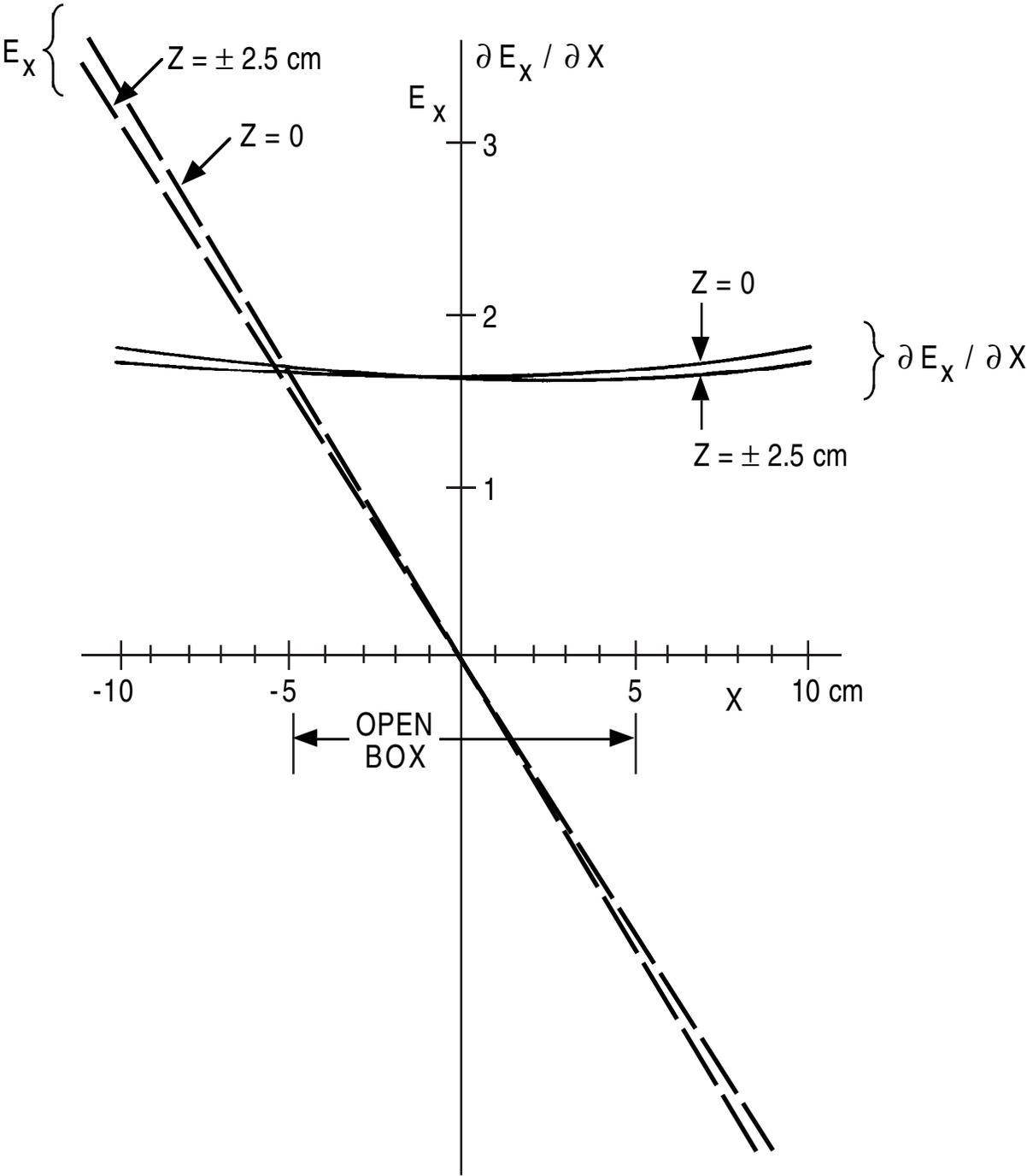
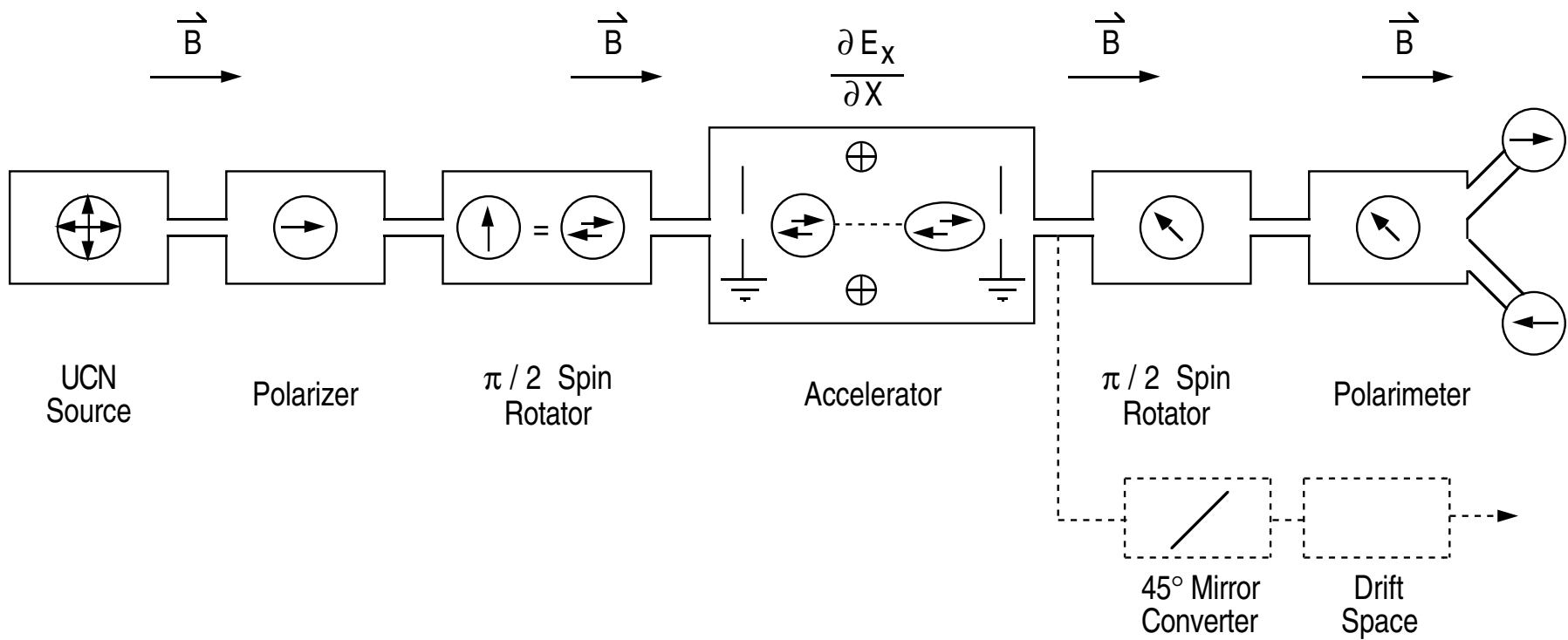
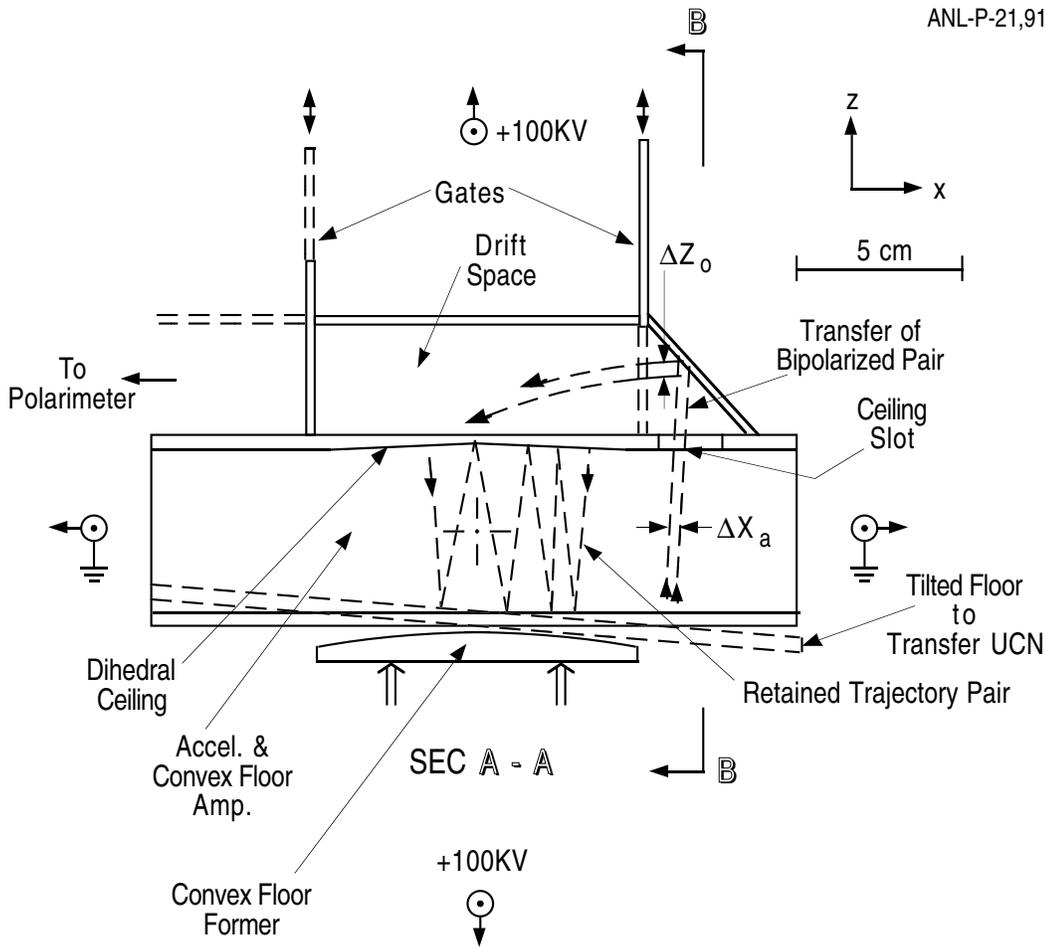
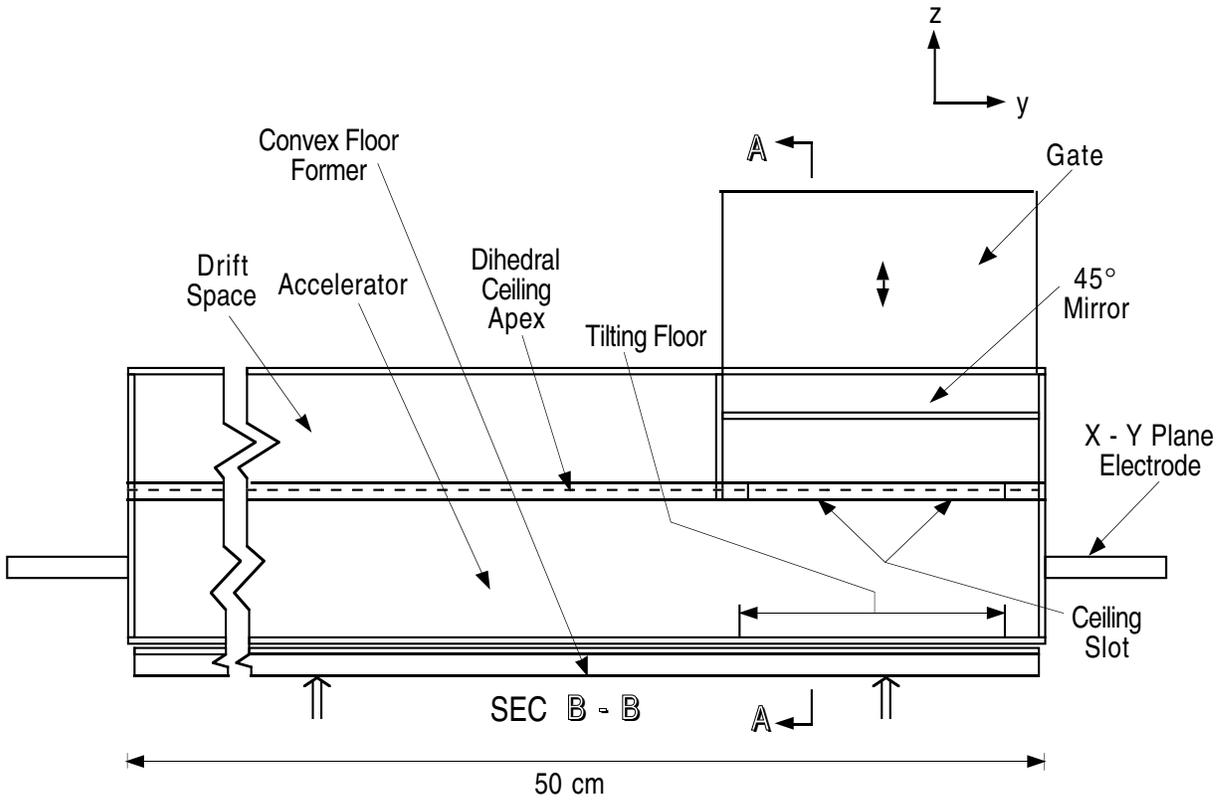
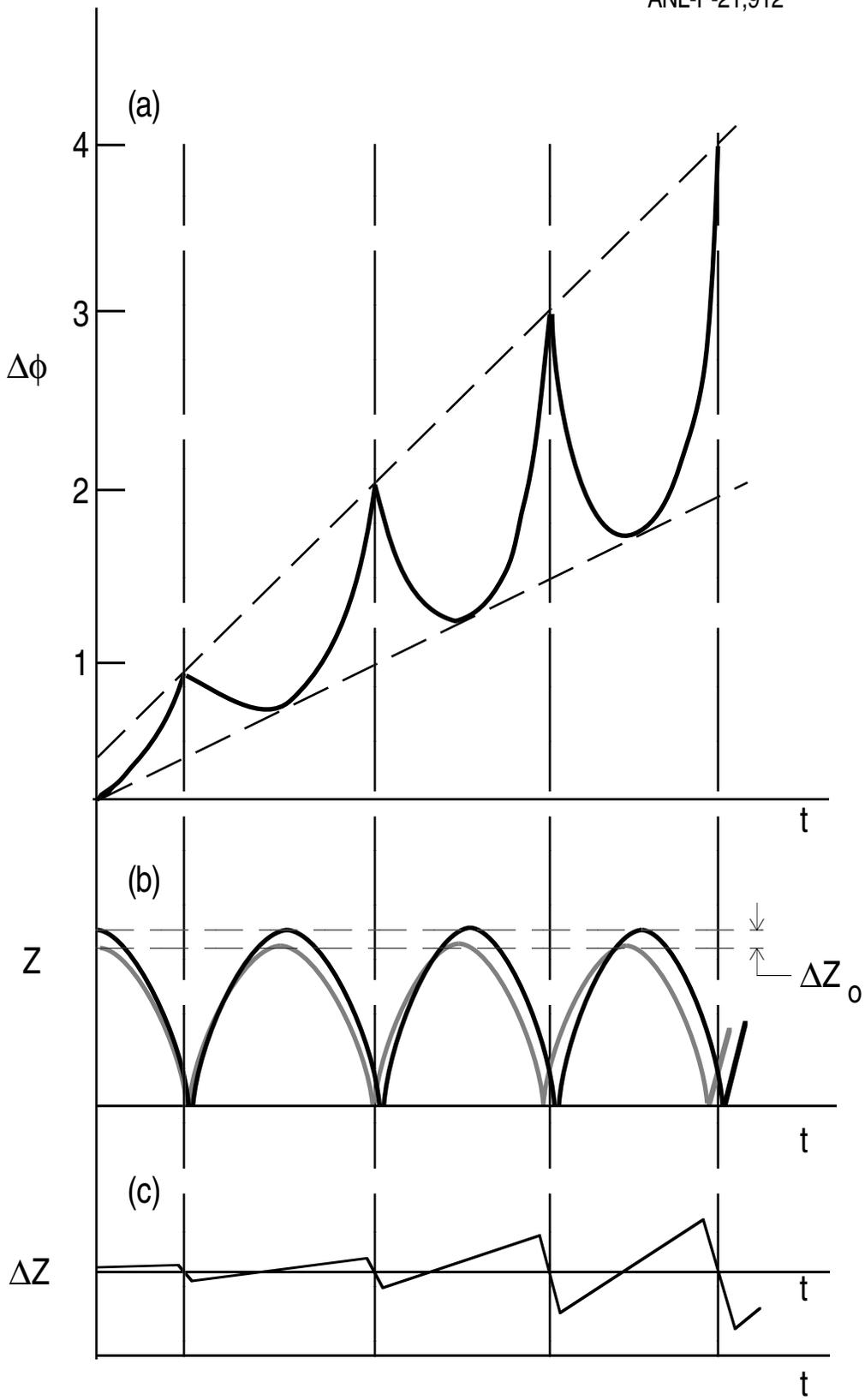
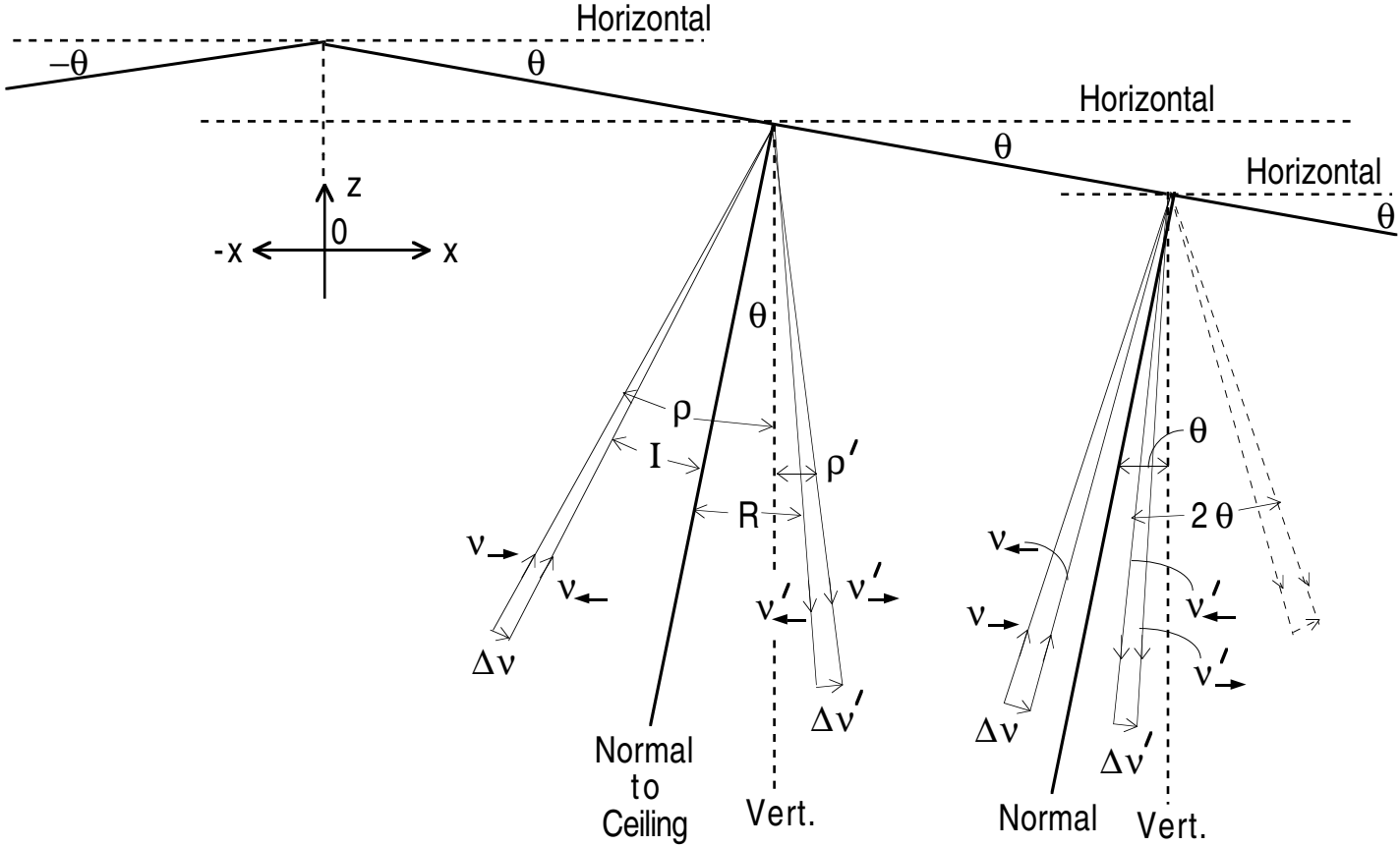
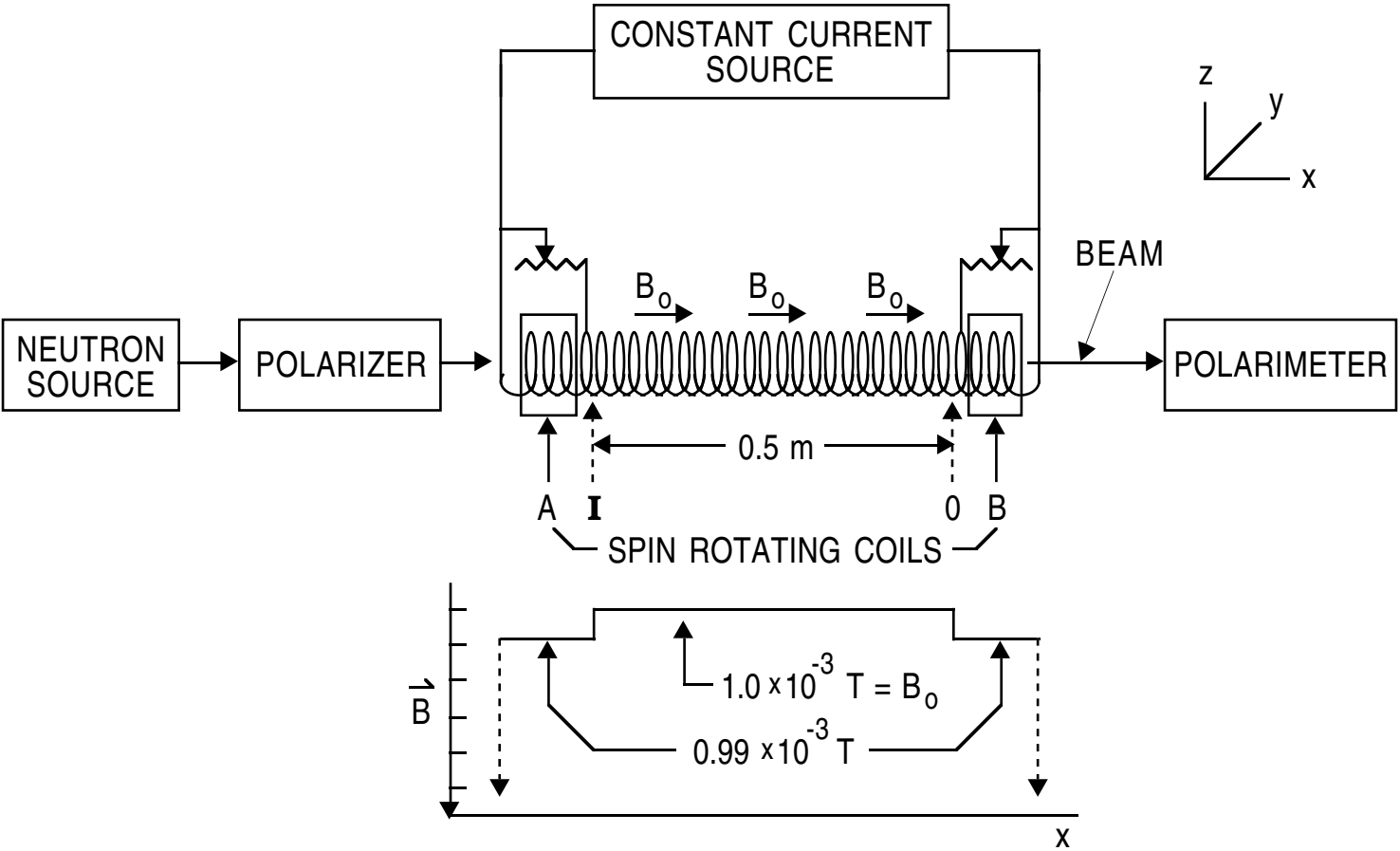
\end{document}